\documentclass[
    ,final            
  ]
  {aipproc}

\layoutstyle{6x9}

\usepackage{epsfig,xspace}
\usepackage{color}

\def\Journal#1#2#3#4{{#1} {\bf #2}, #3 (#4)}

\def\PRL{\em Phys. Rev. Lett.}
\def\PRD{{\em Phys. Rev.} D}

\def\etal {{\em{at al.}}\xspace}

\definecolor{Red}{rgb}{1,0,0}
\definecolor{Blue}{rgb}{0,0,1}
\definecolor{Green}{rgb}{0,1,0}

\def\be{\begin{equation}}
\def\ee{\end{equation}}
\def\bea{\begin{eqnarray}}
\def\eea{\end{eqnarray}}

\def\iab{\ensuremath{\mathrm{ab}^{-1}}\xspace}

\def\yfours{\ensuremath{\Upsilon(\mathrm{4S})}\xspace}

\newcommand{\e}      [1]   {{\ensuremath{\times 10^{{#1}}}}}

\newcommand\tev {\ensuremath{{\mathrm{TeV}}}\xspace}

\usepackage{relsize}
\def\babar{\mbox{\slshape B\kern-0.1em{\smaller A}\kern-0.1em
    B\kern-0.1em{\smaller A\kern-0.2em R}}\xspace}

\begin{document}

\title{\boldmath The SuperB Experiment}

\classification{13.25Hw, 12.39.St, 14.40.Nd} \keywords      {SuperB, New Phyiscs, Lepton Flavor Violation, CP
Violation, Dark Matter}

\author{A.~J.~BEVAN}{
  address={Department of Physics, Queen Mary University of London, Mile End Road, London E1 4NS, England.}
}

\begin{abstract}
In these proceedings I briefly introduce the project prior to discussing the physics programme of the SuperB high
luminosity $e^+e^-$ collider. By measuring the {\em golden matrix} SuperB will be able to elucidate and constrain the
nature of physics beyond the Standard Model.
\end{abstract}

\maketitle


\section{Introduction}

It is widely expected that there is some form of physics beyond the Standard Model (SM) at an energy scale
$\Lambda_{NP} \sim 1 \tev$ generically referred to as New Physics (NP).  The motivation for expecting NP at this scale
is the desire to solve the hierarchy problem without having a finely-tuned model.  In order to balance our
expectations, it is important to consider complementary constraints on NP, for example those imposed by so-called
Flavor Changing Neutral Currents (FCNC). Such constraints typically require $\Lambda_{NP} \geq 10 \tev$. This dichotomy
is known as the (SUSY) flavor problem~\cite{bevan:susyflavour}.  Many models of NP introduce a large number of flavor
changing parameters, for example two-thirds of the parameters of MSSM are related to the flavor-sector, so it is
important for experimentalists and theorists to seriously consider how we can constrain flavor parameters of new
physics.  The energy frontier experiments of the LHC and proposed ILC will directly search for NP, but have a very
limited reach in terms of constraining the flavor sector of the NP~\cite{bevan:hiller}. In order to elucidate the NP
flavor-sector, one needs to have a balanced flavor physics programme, including experiments studying neutrinos, heavy
mesons, and charged leptons. SuperB is an experiment that will perform precision measurements of heavy mesons and
$\tau$ leptons in order to constrain NP scenarios~\cite{bevan:superb,bevan:cdr,bevan:valencia}. The known level of CP
violation in the SM is insufficient to solve the universal asymmetry problem, in other words starting from equal
amounts of matter and antimatter in the Big-Bang we don't understand how the universe has evolved into it's matter
dominated state. The solution to the flavor problem could be related to the solution to the universal asymmetry
problem.

\section{The SuperB experimental programme}

SuperB is a next generation high luminosity asymmetric energy $e^+e^-$ collider proposed to be built near Rome. The
purpose of this experiment is to elucidate the nature of NP. The accelerator will have bunches of 7 GeV electrons
colliding with 4 GeV positrons, with a 15mrad crossing angle.  The Lorentz boost of the resulting center of mass system
in the lab frame will be approximately half that of the current SLAC B-factory. To compensate for the geometrical
effect of the crossing, sextupole magnets will be used before and after the interaction point (IP) in the so-called
{\em crabbed waist} scheme in order to maintain maximal overlap of colliding bunches at the
IP~\cite{bevan:crabbedwaist}. The design luminosity for the accelerator is $10^{36}\, \mathrm{cm}^{-2}
\mathrm{s}^{-1}$, with a total of 75\iab of data being delivered at the $\yfours$ resonance in the first five years of
operation. To achieve this luminosity the vertical beam-size will be of the order of 20nm thus the collider will
operate with a small vertical emittance.  In addition to operating at the $\yfours$ center of mass (CM) energy, the
accelerator will be able to operate at other energies including at the $\psi(3770)$ and $\Upsilon(\mathrm{NS})$
resonances, where $N=1,2,3,$ and 5.

It is anticipated that SuperB will reuse a number of components of the SLAC B-factory, including parts of the PEP-II
accelerator complex, the super-conducting solenoid magnet, barrel of the electromagnetic calorimeter, and the quartz
bars of the particle identification system.  To compensate for the small Lorentz boost of the CM system in the lab
frame, the innermost tracking system will consist of a layer of pixel sensors at a small radius.  This will be
surrounded by a silicon strip vertex detector, and a drift chamber.  Surrounding the tracking system will be a particle
identification system with better response than the \babar DIRC, an electromagnetic calorimeter, and super-conducting
solenoid magnet. The flux return of the solenoid will be instrumented with a system similar to the MINOS scintillating
fibre detector. The SuperB project is described at length in a Conceptual Design Report~\cite{bevan:cdr} and the
proceedings of the 2008 Valencia Physics Workshop~\cite{bevan:valencia}. There is a great deal of international
interest in SuperB and a Technical Design Report (TDR) describing the details of the programme is in preparation. The
TDR is expected to be completed within the two years. The remainder of these proceedings focusses on a few of the
highlights of the NP search potential of SuperB.

Since the discovery of neutrino mixing at the end of the last decade, we know that the SM has an intrinsic level of
charged Lepton Flavor Violation (LFV).  The SM levels of LFV in $\tau$ decay are highly suppressed, however there are
many popular models of NP that can enhance LFV up to $\sim few \e{-8}$.  Such large LFV signals are just beyond the
reach of observation of the existing B-Factories~\cite{bevan:bfactorylfv}, and beyond expectations of LHC
experiments~\cite{bevan:lhclfv}.  One key feature of SuperB is that the beams will be polarized.  The baseline design
incorporates an 80\% polarization of the $e^-$ beam.  Polarized positron beams are more difficult to realize so,
introducing polarized $e^+$ bunches to SuperB is an anticipated upgrade.  The benefit of polarized beams is that the
resulting $\tau$ polarization can be used to suppress backgrounds.  This means that the single event sensitivities of
many of the $\tau$ LFV searches at SuperB are expected to scale with luminosity, rather than by the square root of
luminosity.  The expected sensitivities of SuperB to the LFV golden channels: $\tau\to \mu \gamma$ and $\tau\to 3\mu$
are $2\e{-9}$ and $2\e{-10}$, respectively.  The ratio of the rates of these modes can be used to distinguish between
SUSY and Little Higgs scenarios~\cite{bevan:lfvsusyvshiggs}.  In addition to this, the limit on $\tau\to \mu \gamma$,
when combined with constraints on $\mu \to e\gamma$ from MEG, and $\theta_{13}$ from accelerator and reactor neutrino
experiments can be used to distinguish between NP scenarios~\cite{bevan:lfvdecodingnp}.

There are a number of theoretically clean observables of rare B decays that are sensitive to NP.  For example, the
branching fraction and CP asymmetry of inclusive $b\to s \gamma$ transitions, the branching fraction of $b\to s
\ell^+\ell^-$, the branching fraction of $B^\pm \to \tau^\pm \nu$, and $B^\pm\to K^\pm\nu\overline{\nu}$ are just four
examples of golden modes.  Each mode is a golden channel for one or more NP scenarios.  It is not possible to decipher
the nature of NP by measuring a single mode, however the pattern obtained by measuring the observables sensitive to new
physics will enable us to distinguish between the different scenarios in the literature and understand the NP. SuperB
will be able to measure all of these modes with 75\iab.  In particular with this data sample we will be able to observe
$B^\pm\to K^\pm\nu\overline{\nu}$, and constrain NP up to a \tev energy scale with $B^\pm \to \tau^\pm \nu$.

SuperB will perform a number of precision time-dependent CP asymmetry measurements of loop dominated processes. Several
of these are theoretically clean, such as $B\to \eta^\prime K^0_S$, and $\phi K^0_S$.  These loop processes are quantum
interferometers for high energy new particles.  Any measured deviation from either the Charmonium $\sin2\beta$
measurement~\cite{bevan:sin2beta} or the SM predicted $\sin2\beta$ could be explained by the presence of NP. The
current discrepancies are at the level of $2.1 - 2.7\sigma$~\cite{bevan:soni}.

It is possible to test lepton universality (LU) at SuperB using $\Upsilon(\mathrm{NS})$ decays, where $N=1,2,3$.  Any
departure from universality could be mediated by a number of possible new particles, including a light scalar Higgs,
$A^0$.  The $A^0$ could even be a Dark Matter candidate~\cite{bevan:sanchis}.  Tests of LU have been performed by CLEO
and the B-factories, where the results are consistent with the SM~\cite{bevan:lucleo}.  It will be possible to perform
precision tests at SuperB using data samples one or two orders of magnitude larger than those presently available.
Other exotic models suggest that it is possible to study Dark Matter using decays of light mesons $M$ to invisible
final states. The SM process for such decays is $M\to \nu\overline{\nu}$, and NP could be manifest via decays to
nutrilinos~\cite{bevan:mcelrath}.  The Belle experiment recently searched for $\Upsilon(\mathrm{1S})\to
\nu\overline{\nu}$~\cite{bevan:belle1s}, and have placed non-trivial bounds on Dark Matter as a result.

Recently the B-Factories have observed charm mixing.  It will be possible to perform precision measurements of the
mixing parameters, and to also embark on a programme of searching for different types of CP violation in charm decays
at SuperB.  By running at on the $\psi(3770)$ resonance near charm threshold, the neutral $D$ mesons will be in a
correlated state. Measurements performed using correlated $D$ mesons are analogous to those we are familiar with in the
$B$ meson sector. Thus the study of CP violation and branching fractions in $D$ meson decay will include the
measurement of a number of NP sensitive observables, as well as fundamental tests of the SM and CKM paradigm.

In addition to the aforementioned NP sensitive studies that are possible at SuperB, it will be possible to perform a
number of precision over-constraints of the CKM paradigm.  These could be used to look for higher order contributions
to the CP violation mechanism in nature, and are also essential ingredients to improve the NP search capability of
other flavor experiments: for example studies of $K\to \pi\nu \overline{\nu}$ where theoretical uncertainties are
dominated by lack of knowledge of the CKM matrix~\cite{bevan:augusto}.

\section{Summary}

Most of the parameters required to describe many models of NP are related to flavor couplings. These are almost
completely inaccessible to the energy frontier programme, so to elucidate NP one has to perform a comprehensive set of
measurements constraining flavor changing processes.  SuperB will do just that using hundreds of billions of $B$, $D$,
and $\tau$ decays.  The set of NP sensitive observables and NP scenarios form a {\em golden matrix}.  By measuring
these observables it will be possible to distinguish between the types of NP discussed in the literature, and home in
on the correct description of nature at high energies~\cite{bevan:valencia}. The energy reach of SuperB's NP
constraints can be as high as $\sim 100\tev$. The SuperB community is in the process of preparing a technical design
report.  Once finalized, it is anticipated that it will take five years to construct and commission the SuperB
accelerator and detector complex.  This timescale could lead to data taking as early as 2015, and accumulating 75\iab
of data by 2020. SuperB is described in detail in Refs~\cite{bevan:cdr} and~\cite{bevan:valencia}.


\begin{theacknowledgments}
The author would like to thank the conference organizers for the opportunity to give this talk, and the SuperB
community. This work is supported by the UK Science and Technology Facilities Council, and the Royal Society.
\end{theacknowledgments}



\bibliographystyle{aipproc}   


\begin{thebibliography}{9}

\bibitem{bevan:susyflavour}For example, see Y.~Grossman {\it et al.}, \Journal{Adv.Ser.Direct.High Energy Phys.}{15}{755}{1998}; A.~Masiero and O.~Vives, {\it Particle Physics in the new millenium} {\bf 616} 93 (2003); I~Medeiros and G.~Ross, hep-ph/0612220.
\bibitem{bevan:hiller}G.~Hiller, arXiv:0905.0327.
\bibitem{bevan:superb}\url{http://www.pi.infn.it/SuperB/}.

\bibitem{bevan:cdr}M.~Bona {\it et al.}, arXiv:0709.0451.
\bibitem{bevan:valencia}D.~Hitlin {\it et al.}, arXiv:0810.1312.

\bibitem{bevan:crabbedwaist}P.~Raimondi contribution to La Thuile, March 2009; P. Raimondi {\it et al.}, physics/0702033.

\bibitem{bevan:bfactorylfv} See A.~Cervelli, these proceedings.
\bibitem{bevan:lhclfv}For example see R.~Santinelli and M.~Biasini, {\it CMS NOTE} 2002/037, and K. Mazumdar \Journal{\it
Czechoslov. J. Phys.}{54}{A291}{2004}.
\bibitem{bevan:lfvsusyvshiggs} For example, see Ref.~\cite{bevan:valencia} and references therein.
\bibitem{bevan:lfvdecodingnp} S.~Antusch {\it et al.}, JHEP {\bf 0611} 090 (2006).

\bibitem{bevan:sanchis}M.~A.~Sanchis-Lozano, hep-ph/0610046; M.~A.~Sanchis-Lozano, \Journal{Int. J. Mod. Phys. A}{19}{2183}{2004}.
\bibitem{bevan:lucleo} D.~Besson {\it et al.}, \Journal{\PRL}{98}{052002}{2007}; G.~Bonneaud, proceedings of FPCP 2009.

\bibitem{bevan:mcelrath}B.~McElrath, \Journal{\PRD}{72}{103508}{2005}.
\bibitem{bevan:belle1s}O.Tajima \etal, \Journal{\PRL}{98}{132001}{2007}.

\bibitem{bevan:sin2beta} B.~Aubert {\em et al.}, \Journal{\PRD}{79}{072009}{2009};
  K.~F.~Chen {\em et al.}, \Journal{\PRL}{98}{031802}{2007};
  H.~Sahoo {\em et al.}, \Journal{\PRD}{77}{091103}{2008}.
\bibitem{bevan:soni}E.~Lunghi and A.~Soni, arXiv:0903.5059.

\bibitem{bevan:augusto} For example see A.~Ceccucci, proceedings of Moriond Electroweak, La Thuile (2009).

\end{thebibliography}



\end{document}